\def\b{\begin{equation}}
\def\e{\end{equation}}
\def\ber{\begin{eqnarray}}
\def\eer{\end{eqnarray}}

\def\ie{{\it ie.~\/}}
\def \L {$\Lambda$}

\def \hm1 {$h^{-1}$}
\def\mf{{Minkowski functionals }}

\def\ref{\par\noindent\hangindent\parindent\hangafter1}

\documentclass{iaus}
\usepackage{graphics}

  \checkfont{eurm10}
  \iffontfound
    \IfFileExists{upmath.sty}
      {\typeout{^^JFound AMS Euler Roman fonts on the system,
                   using the 'upmath' package.^^J}%
       \usepackage{upmath}}
      {\typeout{^^JFound AMS Euler Roman fonts on the system, but you
                   dont seem to have the}%
       \typeout{'upmath' package installed. iaus.cls can take advantage
                 of these fonts,^^Jif you use 'upmath' package.^^J}%
      }
  \else
  \fi

  \checkfont{msam10}
  \iffontfound
    \IfFileExists{amssymb.sty}
      {\typeout{^^JFound AMS Symbol fonts on the system, using the
                'amssymb' package.^^J}%
       \usepackage{amssymb}%
         \let\leq=\leqslant
         
      }{}
  \fi

  \IfFileExists{amsbsy.sty}
    {\typeout{^^JFound the 'amsbsy' package on the system, using it.^^J}%
     \usepackage{amsbsy}}
    {}

\newsavebox{\astrutbox}
\sbox{\astrutbox}{\rule[-5pt]{0pt}{20pt}}

\newcommand\etal{\mbox{\textit{et al.}}}

\title{Morphological Statistics of the Cosmic Web}
\author{Sergei F.Shandarin}
\affiliation{
   Department of Physics and Astronomy, University of Kansas,KS
  66045, USA\\ email: sergei@ku.edu}

\pubyear{2004}
\volume{195}
\pagerange{1--8}
\date{?? and in revised form ??}
\jname{Outskirts of Galaxy Clusters: intense life in the suburbs}
\editors{A. Diaferio, ed.}
\begin{document}

\maketitle

\begin{abstract}
 We report the {\em first} systematic study of the supercluster-void
network in the $\Lambda$CDM concordance cosmology treating voids 
and superclusters on an equal footing. We study the dark matter density
field in real space smoothed with the $L_s = 5$ \hm1 Mpc Gaussian
window.
Superclusters and voids are defined as individual 
members of over-dense and under-dense
excursion sets respectively.
We determine the morphological properties of the 
cosmic web at a large number of dark matter density levels 
by computing Minkowski 
functionals for every supercluster and void.
At the adopted smoothing scale individual superclusters 
totally occupy no more than about 5\% of
the total volume and contain no more than 20\% of mass if the largest
supercluster is excluded. Likewise,
individual voids totally occupy no more than 14\%  of volume
and contain no more than 4\% of mass if the largest void is excluded.
 The genus of individual superclusters can 
be $\sim 5$ while 
the genus of individual voids reaches $\sim 55$, implying significant
amount of substructure in superclusters and especially in voids.
Large voids are typically distinctly non-spherical. 
\end{abstract}
\firstsection
\section{\bf Introduction}
Redshift galaxy catalogues reveal a universe permeated by an interpenetrating 
network of superclusters and voids.
It therefore becomes important to understand and quantify the
geometrical and topological properties of 
large scale structure in an \L CDM cosmology in a deep
and integrated manner.
The main aim of this talk is to study
the supercluster-void
network in \L CDM cosmology with emphasis on the sizes, shapes and topologies
of individual superclusters and voids.
Concretely, we study the geometry and topology of isodensity surfaces 
$\delta({\bf x})\equiv \delta \rho({\bf x})/\bar{\rho}=const$.
At a given threshold $\delta_{th}$ regions having 
higher then threshold density ($\delta > \delta_{th}$) will be called
``superclusters'', while 
regions with $\delta < \delta_{th}$ will be 
called ``voids'' 
We employ an elaborate surface
modeling scheme, SURFGEN (short for `surface generator'), that allows us to
determine the geometry, morphology and topology of excursion sets
defined on a density field in a very comprehensive manner
(Sheth \etal 2003). Working with the density field
also permits us to determine the morphological properties of the {\em
full excursion set} describing the supercluster-void network. More
detailed information is then gleaned at one particular threshold
(usually associated with percolation) at which shapes and sizes of individual
superclusters and voids yield rich information about properties of the
cosmic web to which we belong.

We use dark matter
distributions in a flat model with $\Omega_0$ = 0.3,
$\Omega_{\Lambda}$ = 0.7, $h=0.7$ 
(\L CDM). The initial spectrum was taken with the 
shape parameter  $\Gamma = 0.21$.
The amplitude ($\sigma_8=0.9$) of the power spectrum in the model is
set so as to reproduce the observed abundance of rich galaxy clusters
at the present epoch  (see Jenkins \etal 1998 for details).
The study of mocked galaxy distributions was done by Sheth (2003).

SURFGEN operates on three-dimensional pixelized maps. Therefore we
first generate the density field from the distribution of dark matter
particles.  This process was described in detail in Sheth \etal (2003); 
here we present a brief summary.  The data
consist of $256^3$ particles in a box of size 239.5 \hm1 Mpc. We fit a
$128^3$ grid to the box.  
We follow the smoothing technique used by Springel \etal (1998)
which they adopted for their preliminary topological analysis of the
Virgo simulations.  We study the field smoothed with $L_s = 5$ \hm1 Mpc
which is a fiducial
smoothing scale in many studies of both density fields in N-body
simulations and galaxy fields from redshift surveys; see for example, 
Grogin \& Geller (2000).  The Gaussian kernel for smoothing that we adopt
here is 
$ W(r) = \pi ^{-3/2} L_s^{-3} exp{\left( -r^2/L_s^2\right)}  $.
Since the kernel is isotropic and uniform , it is likely to diminish the true
extent of anisotropy in filaments, pancakes and voids.  
This effect could be
minimized by considering adaptive kernels or smoothing
techniques based on the wavelet transform.  An even more ambitious
approach is to reconstruct density fields using Delaunay tesselations
using a technique reported by van de Weygaert (2002). 
 
Four Minkowski functionals are effective non-parametric descriptors of
the morphological properties of surfaces in three dimensions
Mecke \etal (1994). They are:
{\it Volume} $V$ enclosed by the surface, $S$;
{\it Area} $A$ of the surface;
{\it Integrated mean curvature} $C$ of the surface 
$
C = \frac{1}{2}\oint_S{\left({1\over R_1} + {1\over R_2}\right)da}, 
$
where $R_1$ and $R_2$ are the principal radii of curvature at a given
point on the surface; and finally,
{\it the Euler characteristic}
$
\label{eq:euler} \chi = \frac{1}{2\pi}\oint_S{\left({1\over
      R_1R_2}\right)da}.  
$
The genus is uniquely related
to the Euler characteristic $G = 1 - \chi/2$ and thus carries exactly
the same information. We measure the above parameters for every region
in both over-dense and under-dense excursion sets at 99 density
thresholds equispaced in the filling factor from $FF=0.01$ to
$FF=0.99$ where $FF$ is the fraction of the total volume occupied by the
excursion set.
As demonstrated in Sahni \etal (1998) and Sathyaprakash \etal (1998) 
particular ratios of Minkowski
functionals called ``Shapefinders'' provide us with a set of
non-parametric measures of sizes and shapes of objects.  Therefore, in
addition to determining MFs we shall also derive three quantities having
the dimensions of length that can be associated with three
characteristic sizes:  $T$
(Thickness), $B$ (Breadth) and $L$ (Length) defined as follows:
$
T = 3V/A, \ \  B = A/C, \ \ L = C/4\pi.
$ 
The three Shapefinders describing an individual region bounded by one
or several isolated surfaces of constant density have dimensions of
length and provide us with an estimate of the regions extensions.   
The choice of the coefficients in 
the above equations results in a sphere having all three sizes equal to its
radius $T=B=L=R$.  It is worth
noting that $T$, $B$ and $L$ are only the estimates of three basic
sizes (semi-axes) of an object which work quite well on such objects
as a triaxial ellipsoid and torus.
An indicator of `shape' is provided by a pair of dimensionless
Shapefinder statistics
$
P = (B-T)/(B+T);~~~ F = (L-B)/(L+B), 
$
where $P$ and $F$ are measures of Planarity and Filamentarity
respectively ($P, F \leq 1$).  A sphere has $P = F = 0$, an ideal
filament has $P = 0, F = 1$ while $P = 1, F = 0$ for an ideal pancake.
\section{Results}
\begin{figure}
 {\resizebox{6.5cm}{6.5cm}{\includegraphics{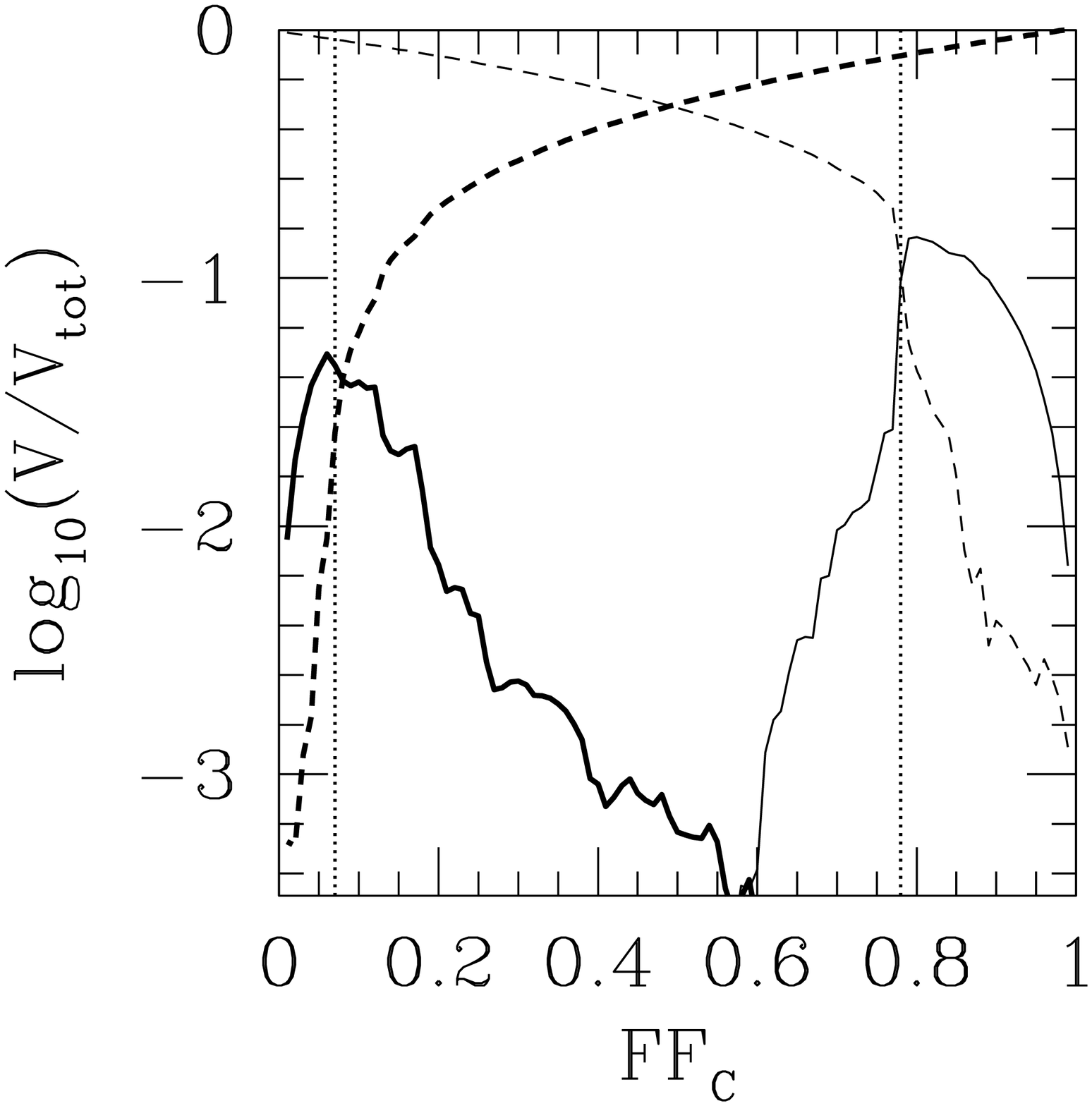}}}
 {\resizebox{6.5cm}{6.5cm}{\includegraphics{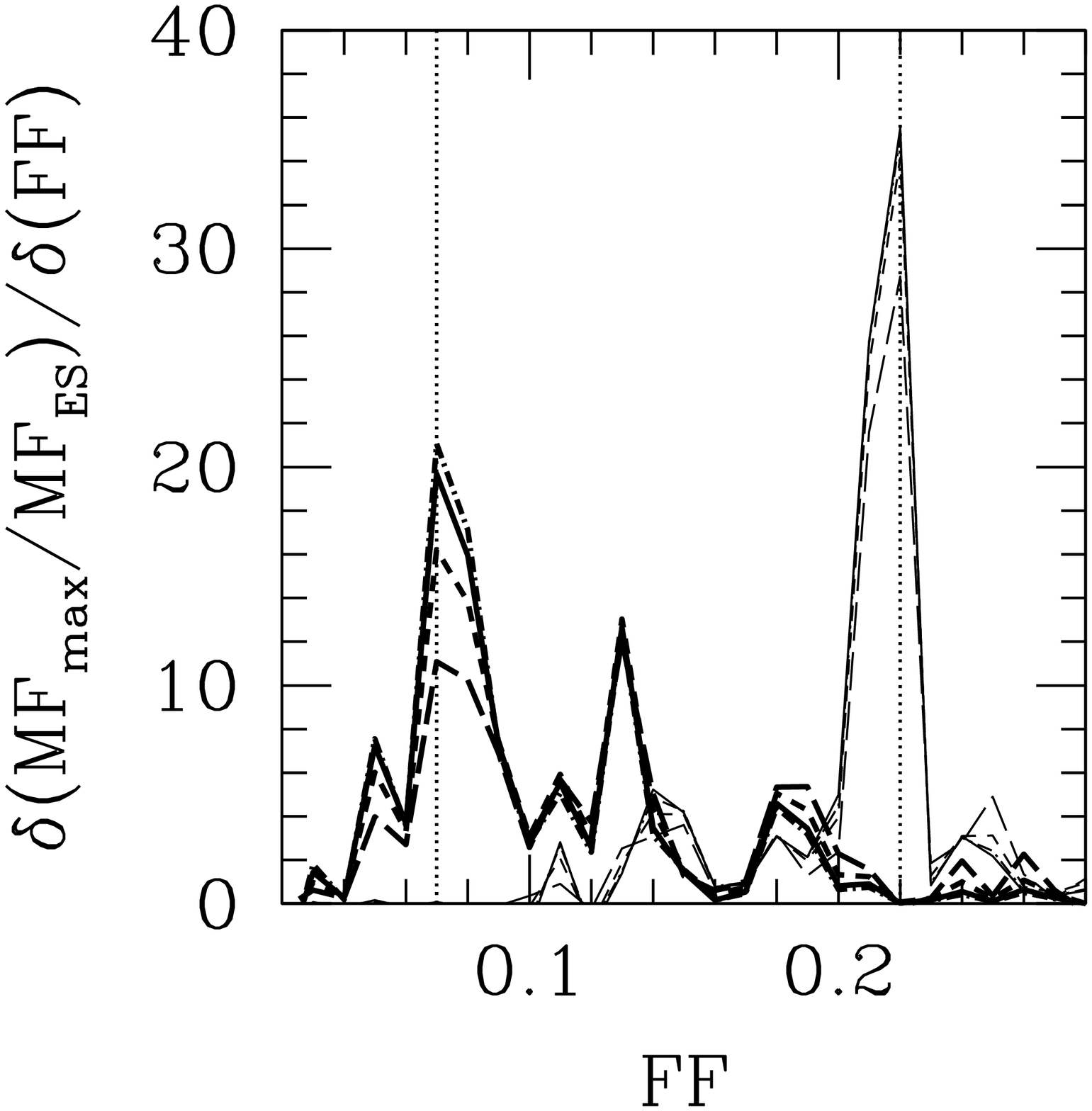}}}
  \caption{{\small Left: the fractions of the total volume occupied by 
the largest supercluster (thick dashed line), all superclusters 
but the largest one (thick solid line), 
largest void (thin dashed line), and all voids but the largest one 
(thin solid line)
are shown for the density field in the \L CDM model smoothed 
with  $L_s$=5 \hm1 Mpc  as  a function of  the filling factor, $FF_C$.
Right: the estimates of the percolation thresholds.
Thick lines show the rate of growth $\delta m^{(i)}/\delta (FF)$ 
for four estimators (Shandarin \etal 2003) for superclusters 
as a function of $FF_C$. All four curves consistently peak at 
$FF_C = 0.07$.
Thin lines show similar quantities for voids with a distinct peak at 
$FF_C = 0.22$.  Vertical dotted lines mark the percolation 
thresholds.}}
\end{figure}
\begin{figure}
 {\resizebox{6.cm}{6.cm}{\includegraphics{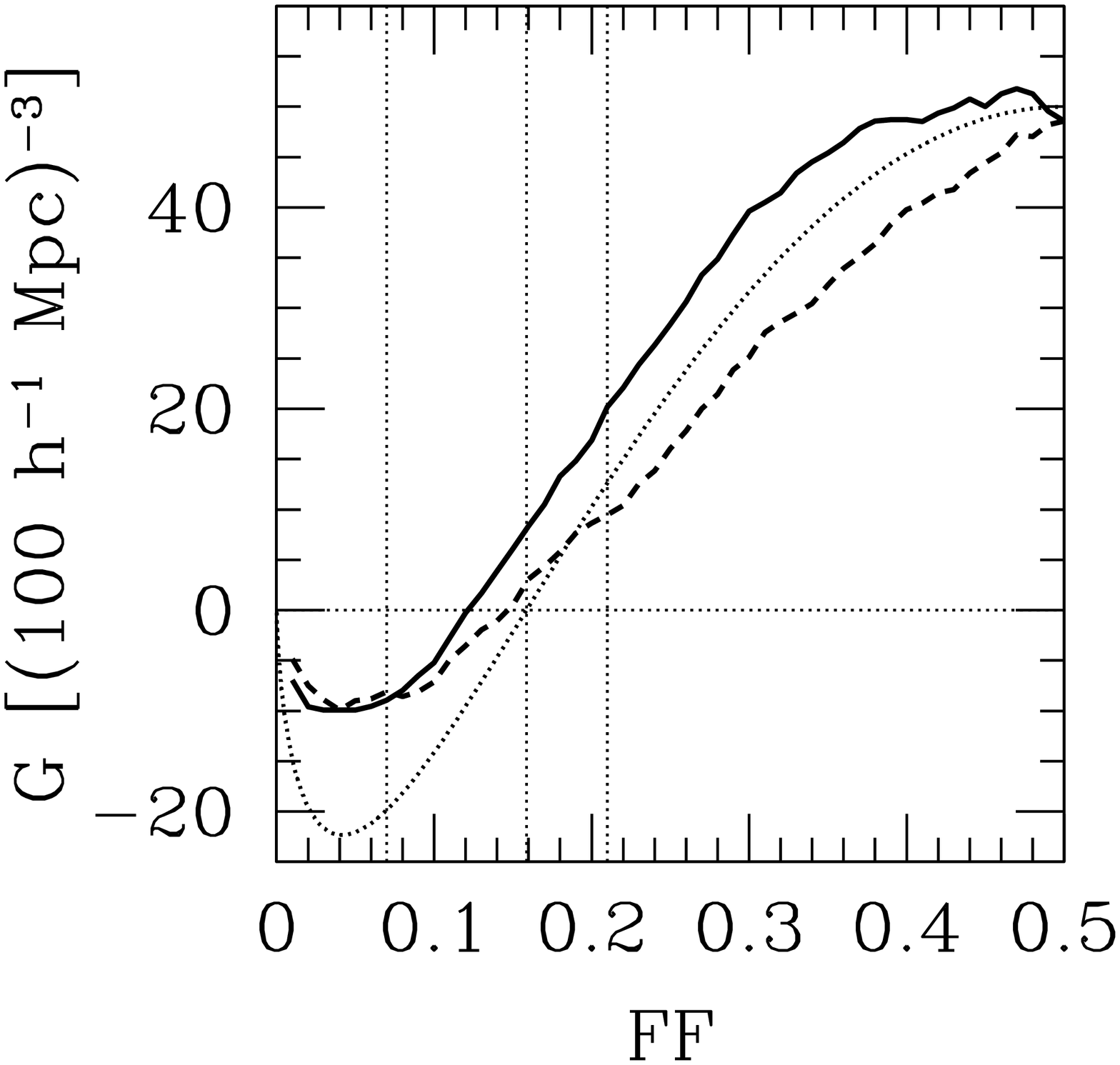}}}
 {\resizebox{6.cm}{6.cm}{\includegraphics{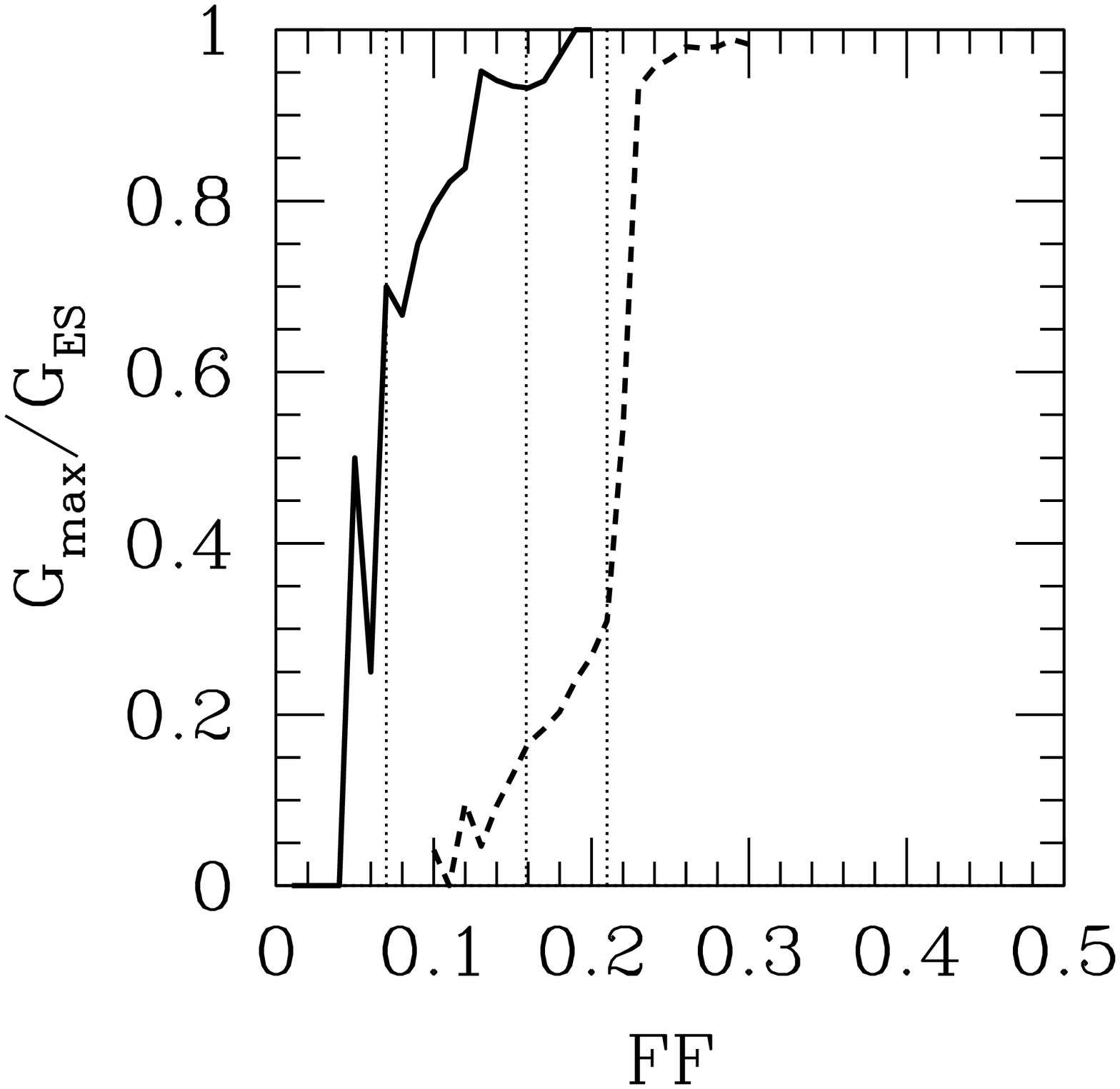}}}
  \caption{{\small Left: the global genus is shown as a function of
the filling factor for the density field smoothed with $L_s=5$ $h^{-1}$ Mpc.
The half of the curve corresponding to high density thresholds is 
plotted as a function of $FF_C$ (solid line) while the other half 
corresponding to low density thresholds is plotted as a function of $FF_V$
(dashed line).  For comparison, the dotted line shows the Gaussian genus 
curve having the  same amplitude.  
Right: the percolation transitions in the same density field
as indicated by the genus of the largest  supercluster
(solid line) and largest void (dashed line). The vertical dotted lines 
mark two percolation thresholds in the $\Lambda$CDM
($FF \approx 0.07$ and $FF \approx 0.22$)
and Gaussian field ($FF \approx 0.16$) in both panels.
}}
\end{figure}
Understanding percolation is essential for understanding the morphology of
the supercluster -- void network. Percolation is important because the
properties of superclusters and voids radically change at the
percolation transitions (see Fig. 1). 
At relatively
high thresholds $\delta _C > 1.8$ corresponding to small filling
factors $FF_C < 0.07$ the largest supercluster has insignificant
volume and mass compared to the total volume or mass contained in the
over-dense excursion set, $\delta > \delta _{\rm TH}$.  During the
percolation transition at $FF_C \approx 0.07$ corresponding to
$\delta _C \approx 1.8$, both volume and  mass in the largest
supercluster rapidly grow, overtaking the volume and mass in the entire
excursion set, and completely dominating the entire sample from 
this point onwards.
\begin{figure}
 {\resizebox{6.5cm}{6.5cm}{\includegraphics{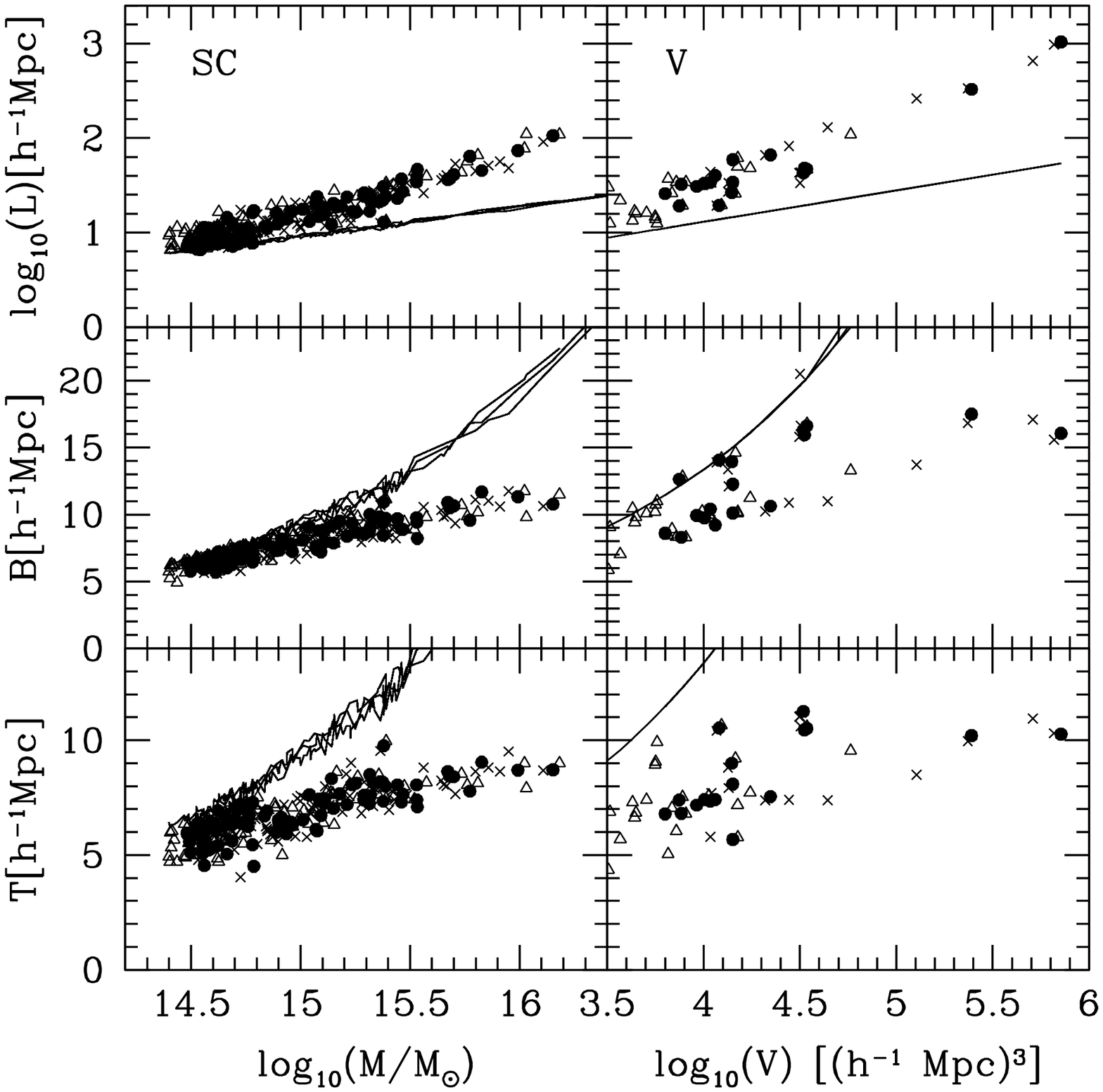}}}
 {\resizebox{6.5cm}{6.5cm}{\includegraphics{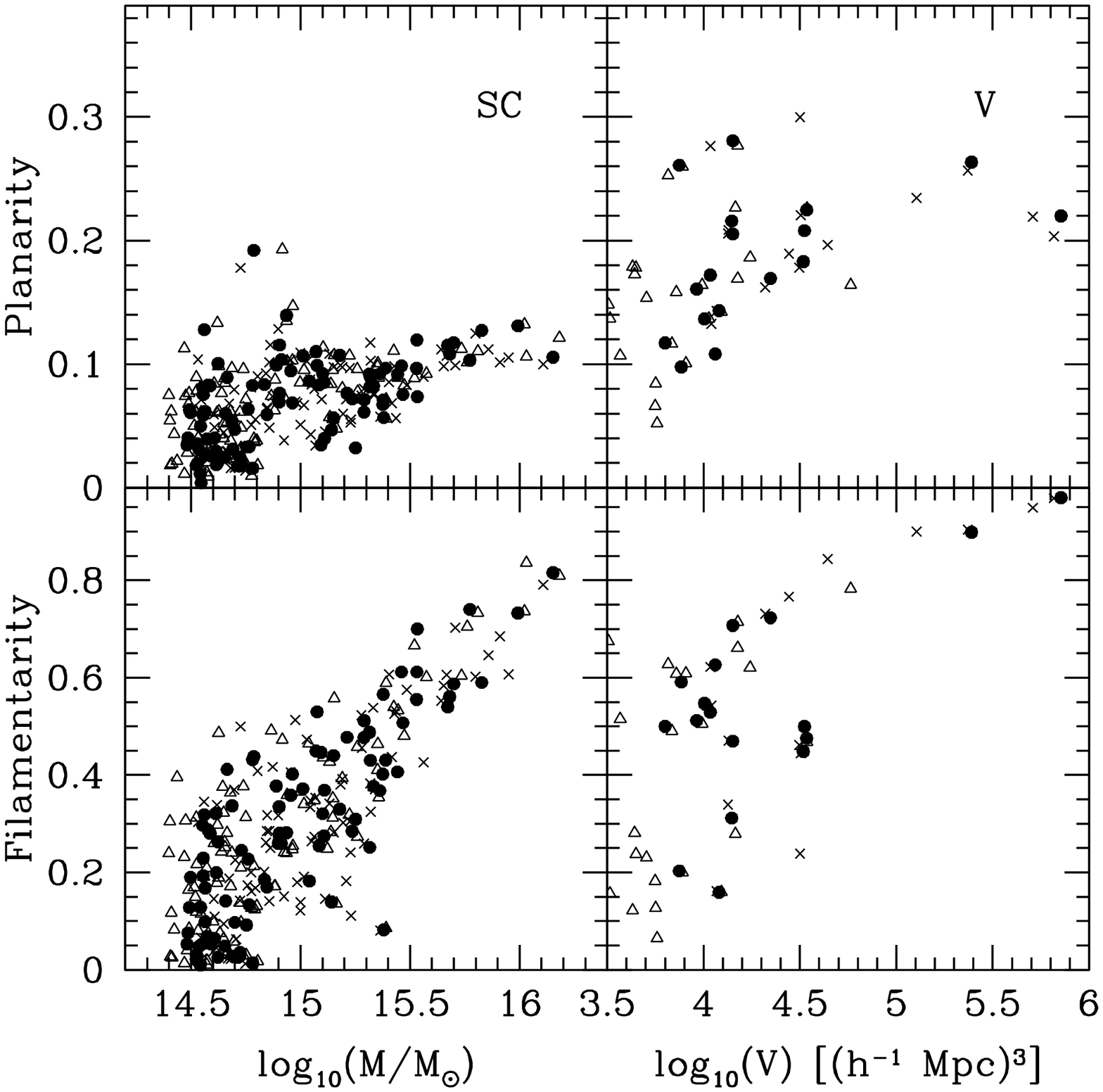}}}
  \caption{{\small Left: length, breadth, and thickness versus mass 
for superclusters
and versus volume for voids. Solid circles show the relation
at percolation thresholds: $FF_C=0.07$ for superclusters and $FF_V=0.22$ 
for voids. 
Crosses show the parameters before percolation ($FF_C=0.06$ 
for superclusters and $FF_V=0.21$ for voids) and empty triangles 
after percolation ($FF_C=0.08$ for superclusters and $FF_V=0.23$ for voids).
Solid lines show the radius of the sphere having the same volume as 
the supercluster or void. 
Right: planarity and filamentarity vs mass (for superclusters)
and vs volume (for voids) at percolation.
Notations are as in the left panel
}}
\end{figure}
\begin{figure}
 {\resizebox{6.5cm}{6.5cm}{\includegraphics{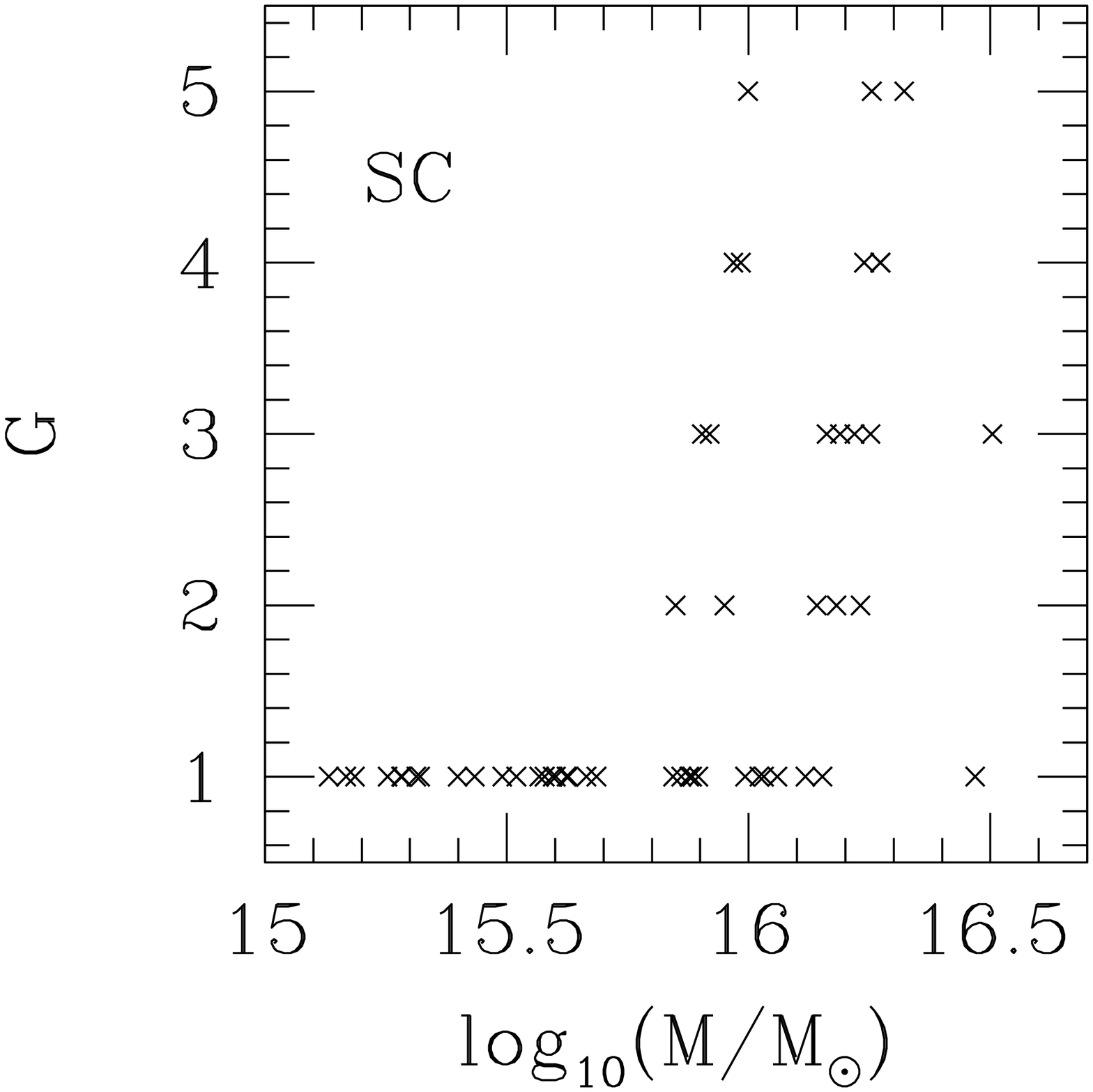}}}
 {\resizebox{6.5cm}{6.5cm}{\includegraphics{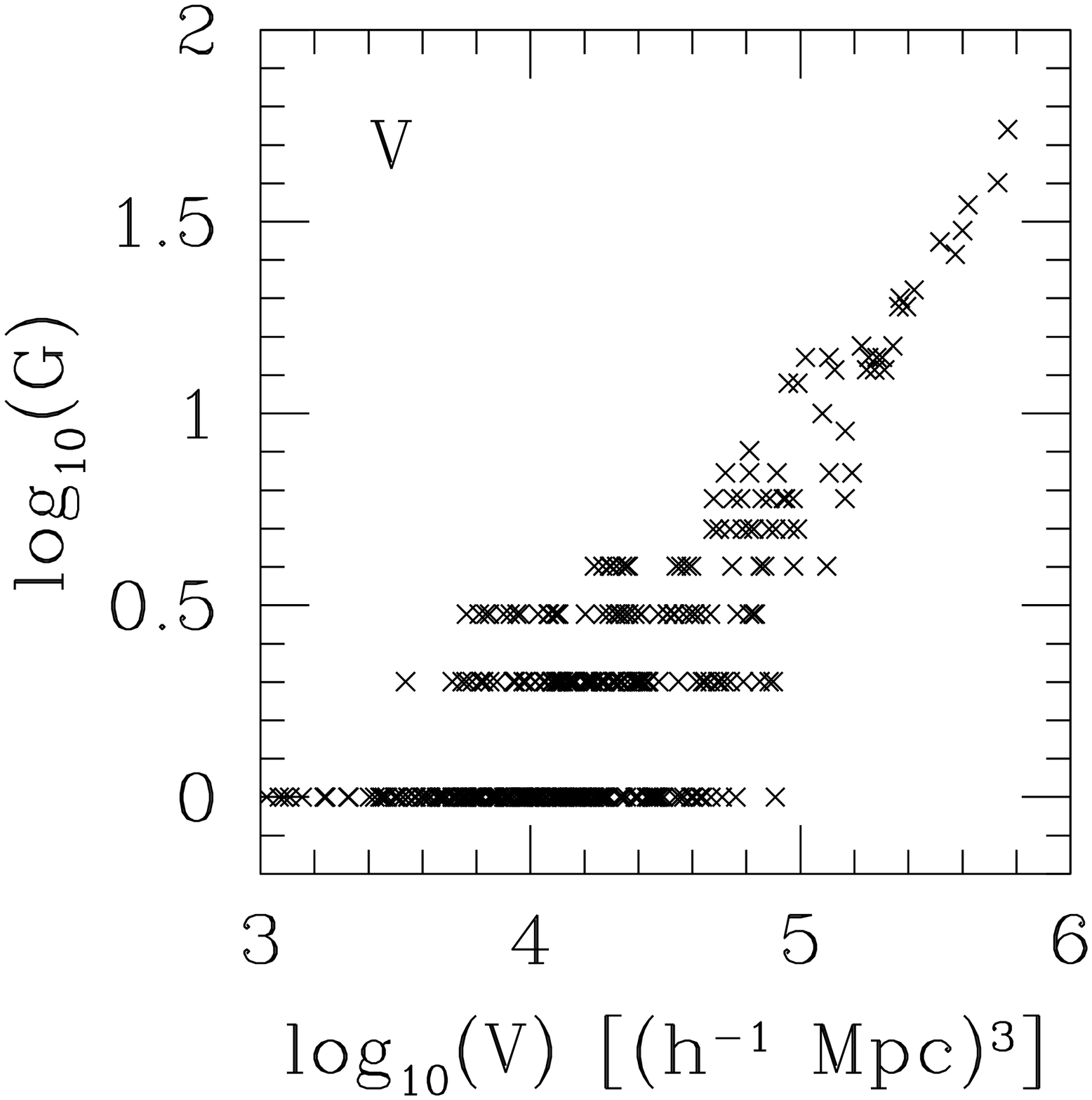}}}
  \caption{{\small Left: genus-mass relation for supercluscters.
Right: genus-volume relation for voids.
Every isolated supercluster and void having
genus greater than zero at all density thresholds is shown.}}
\end{figure}
The largest void behaves in a qualitatively similar manner if plotted
versus $FF_V$. At $FF_V < 0.22$, $\delta _V < -0.5$ its volume
is small compared to the volume of the under-dense excursion set,
$\delta < \delta_{\rm TH}$, but at the percolation transition $FF_V
\approx 0.22, \delta _V \approx -0.5$, it takes over and from then 
on remains the
dominant structure in the under-dense excursion set.  Since $FF_C
\equiv 1-FF_V$ the void percolation transition takes place at $FF_C
\approx 0.78$ as shown in Fig. 1 (left panel).
Two obvious conclusions can be drawn from the above discussion.  First, at
percolation the object having the largest volume becomes very
different from all remaining objects, therefore it must be studied separately.
Second, individual objects -- both superclusters and voids -- must be
studied in the corresponding phase {\em before percolation occurs} in the
corresponding phase. Both superclusters and voids reach
their largest sizes, volumes and masses just before percolation sets in.
The percoation thresholds can be accurately measured from the peaks
in the rate of growth of the MFs of the largest supercluster or void
as a function of $FF$ (Fig. 1 right panel).

It is interesting to compare the percolation and genus statistics.
Both were suggested as tests for the connectedness of 
the large-scale structure.  First, Zel'dovich and Shandarin
(Zel'dovich 1982, Shandarin 1983, Shandarin \& Zel'dovich 1983) 
raised the question of topology of large-scale
structure and suggested percolation statistic as a discriminator
between models.  The percolation test was first applied to a redshift catalog
compiled by J. Huchra and Rood (Zel'dovich \etal 1982; 
Einasto \etal 1984)
who found that the connectedness in this catalog was
significantly stronger than for a Poisson distribution. In contrast, a
non-dynamical model having approximately correct
correlation functions up to the fourth order (Soneira \& Peebles 1978)
showed significantly weaker connectedness than in the observed catalog. 
Thus, percolation was able to detect connectedness in the galaxy distribution.
It was also demonstrated that three lowest order correlation functions  
(two-, three- and four-point functions) are not sufficient to detect
the connectedness in the galaxy distribution. Figure 2 illustrate
the difference between the global genus (Fig. 2 left panel) and percolation
staitistic (Fig. 1 left panel and Fig. 2 right panel): 
at the percolation threshold many properties of the excursion set 
significantly change while the global genus curve remain smooth and
featureless.

Three characteristic sizes and shapes of superclusters and voids can
be estimated from \mf.  
It is surprising that the thickness of
superclusters depends on the threshold quite  weakly; it is within 4$-$6
h$^{-1}$Mpc interval for a range of thresholds between $0 < \delta
< 6$.  This may indicate that the actual thickness of
superclusters is significantly smaller and the measured values reflect
the width of the smoothing window. The breadth of superclusters is not
much greater than thickness and it is likely that it is also
affected by the
width of the filtering window.  Voids are a little fatter than superclusters
and their thickness reaches about 9 h$^{-1}$ Mpc at the percolation
threshold.  Voids are also 
wider and longer than superclusters.
Recalling that the size parameters are normalized to the radius of the
sphere rather than to diameter we conclude that the longest 25\% of
superclusters are longer than about 50 h$^{-1}$ Mpc and 25\% of voids are
longer than about 60 h$^{-1}$ Mpc.

The correlation of the sizes ($L$, $B$, $T$) and shape parameters ($P$, $F$)
of superclusters and voids  with the masses (superclusters) and volumes 
(voids)  are shown in Fig. 3. It was expected that all 
three sizes of the superclusters  may
correlate with their masses and the sizes of the voids correlate with
their volumes. It is perhaps also not surprising that the filamentarity of 
superclusters increases with the mass.
However, the strong correlation of the filamentarity  ($F$) 
of voids with their volumes is quite unexpected and may require 
reconsidering of some theoretical models of the void evolution.
Figure 4 (left panel) demonstrates that the palnarity of superclusters 
is quite small that indicates that pancakes are not typical
structures in the $LCDM$ cosmology.
In addition, we have found a very strong indication of complex 
geometry and nontrivial topology in largest superclusters (Fig. 4 left panel) 
and especially in voids (Fig. 4 right panel).
\section{Summary}
  Individual superclusters totally occupy no more than about 5\%
  of the total volume and comprise no more than 20\% of mass if the
  largest (\ie percolating) supercluster is excluded. 
  The maximum of the total volume and mass comprised by all 
superclusters except 
  the largest one is reached approximately at
  the percolation threshold: $\delta \approx 1.8$ corresponding to
  $FF_C \approx 0.07$.
  Individual voids totally occupy no more than 14\% of volume and
  contain no more than 4\% of mass if the largest void is excluded.
  The maximum of the total volume and mass comprised by all voids
  except the largest one is reached at about the void
  percolation threshold: $\delta \approx -0.5$ corresponding to
  $FF_V \approx 0.22$.
  Between these two percolation thresholds all superclusters and voids
  except the largest ones take up no more than about 10\% of volume and
  mass. Both largest supercluster and void span throughout the whole
  space and have a very large genus. Therefore neither has well defined 
  sizes, volumes, masses or easily defined shapes.\\
\indent     The sizes of voids are significantly larger than those of 
  superclusters even in the density field smoothed with $L_s = 5\ h^{-1}\ Mpc$.
  The length of a quarter of the most massive superclusters 
 exceeds 50 h$^{-1}$ Mpc. The most voluminous voids are even longer: 
25\% of them are longer than 60 h$^{-1}$ Mpc. The longest 
non-percolating supercluster
is as long as 100 h$^{-1}$ Mpc and the longest non-percolating
void is as long as  200 h$^{-1}$ Mpc.
Both are comparable to the size of the box ($239.5\ h^{-1}\ Mpc$)
and therefore may be affected by the boundaries.\\
\indent    The genus value of individual superclusters can be $\sim 5$ while
the genus of individual voids can reach $\sim 55$ (Fig. 4). 
This implies significant
amount of substructure in superclusters and especially in voids.
This is in a general agreement with other studies of voids
(Kofman \etal 1992; Peebles 2001; Gottl\"{o}ber \etal 2003).\\
\indent   One of our main results is that voids, as defined through the density
field can be distinctly non-spherical.
Whether this result carries over to voids in galaxy surveys will
depend upon the nature of the baryon-dark matter biasing and also on
whether the density field is sampled in real or in redshift space.\\
\indent   The percolation thresholds  as well as some other parameters
depend on the smoothing scale
and for smaller smoothing scales or adaptive filtering windows 
the supercluster percolation threshold must decrease ($FF_C^{perc.} < 0.07$)
and the void percolation threshold increase ($FF_V^{perc.} > 0.22$).
\begin{acknowledgments}
I am thankful to Antonaldo Diaferio for careful reading of 
the manuscript and useful critical comments.
\end{acknowledgments}

\end{document}